\documentclass{article}

\usepackage[margin=4cm]{geometry}
\usepackage{amsfonts}
\usepackage{amsmath}
\usepackage{graphicx}
\usepackage{mathrsfs}
\usepackage{bm}
\usepackage{amsbsy,amsmath}
\usepackage[colorlinks=true,citecolor=black,linkcolor=black,backref=true]{hyperref}   

\newcommand{\ket}[1]{\left\vert #1 \right\rangle}

\begin{document}

\title{\textbf{Noncovariant gauge fixing in the quantum Dirac field theory of atoms and molecules}}
\author{\normalsize{A. Stokes}}

\date{\normalsize{\today}}

\maketitle
 
\abstract{The formalism of quantum mechanical gauge fixing in quantum electrodynamics (QED) is extended using techniques from non-relativistic QED. This involves expressing the redundant gauge degrees of freedom through an arbitrary functional of the gauge invariant transverse degrees of freedom. Particular choices of functional can be made to yield the Coulomb or Poincar\'{e} gauge representation of the Hamiltonian. The Hamiltonian we derive therefore serves as a good starting point for the relativistic description of atoms and molecules. Important Implications of the gauge freedom present in the Hamiltonian with regards to the ontology of QED in general are discussed.}

\section{Introduction}

The quantum description of atoms, molecules and optical radiation is traditionally based on the Coulomb gauge or Poincar\'{e} formulations of non-relativistic QED \cite{power1,loudon1,craig,cohen1,cohen2}. An obvious disadvantage of such an approach is the non-relativistic treatment, which to be as accurate as possible should be replaced with a relativistic treatment involving a quantized Dirac rather than Schr\"odinger matter field. 

A second drawback is the restriction from the outset to a particular gauge, from which it is difficult to explore the full implications of the gauge freedom present in the theory. Furthermore, though the Coulomb gauge and Poincar\'{e} gauges are predominant in the literature certain admixtures of these gauges have proven useful in quantum optics indicating the potential benefit of a broader, unifying formulation \cite{drummond,babiker,stokes}. The aim of this paper is to address both of these issues. First using a quantized Dirac field description for the material degrees of freedom and second keeping at the forefront the gauge freedom present in the theory.

The latter is achieved following ref.\cite{woolley1} by re-expressing the gauge dependent longitudinal degrees of freedom using the gradient of a linear functional of the gauge invariant transverse degrees of freedom. The functional is defined in terms of an arbitrary $c$-number function ${\bm g}({\bf x},{\bf x}')$, which then carries the gauge freedom of the theory.

As a description of the Maxwell field we use a Hilbert space of Schr\"odinger wave functionals of the vector potential in which an inner product can be defined through functional integration \cite{jackiw}. The material degrees of freedom can be defined similarly in terms of a Hilbert space of functionals of a Grassman field \cite{jackiw}.

To obtain a Hamiltonian in an arbitrary gauge (defined by ${\bm g}$), which is fully relativistic in the material degrees of freedom we combine these elements with an adaptation of the quantum mechanical gauge fixing method presented in \cite{lenz}. In this approach one initially adopts the Weyl gauge whereby the scalar potential is set equal to zero. From there the physical subspace of states (wave functionals) is defined as the subspace of states vanishing under the action of the Gauss law constraint. The Hamiltonian can be found in different gauges through the use of unitary gauge fixing transformations acting on the Weyl gauge Hamiltonian. 

Particular choices of the function ${\bm g}$ can be made subsequently to render the Hamiltonian in a fixed gauge. Two particular choices yield the Coulomb gauge and Poincar\'e gauge Hamiltonians.

On the one hand our results are of importance for the relativistic theory of atoms and molecules, and on the other they provide an interesting extension of the quantum mechanical gauge fixing formalism already employed in relativistic QED. We will show that this extension has important implications with regards to the ontology of QED in general, because it allows for a thorough exploration of the gauge freedom of the theory in canonical (Hamiltonian) form. The requirement of gauge invariance of a result is translated into the requirement that it be independent of the choice of the arbitrary function ${\bm g}$.

There are four sections to this paper. In section 2 we use the quantum mechanical gauge fixing formalism to obtain a relativistic Hamiltonian in an arbitrary gauge. In section 3 we discuss the implications of the gauge freedom still present in the formulation. We then define useful gauge invariant operators and as an application address the problem of causality in spatially separated material systems \cite{fermi}. In section four we finish with a brief conclusion of our results.

\section{The QED Hamiltonian in an arbitrary gauge}\label{2}

We will begin with a subsection outlining how the general idea of gauge fixing we are going to employ works in the simple case of classical electrodynamics. The aim is then to extend this idea to relativistic QED. To this end we start the following subsection formally with the QED Lagrangian and obtain from it the Hamiltonian and canonical operators, which are supposed to satisfy canonical commutation relations. We go on to identify the constraint Gauss' law, which defines the physical subspace of states and review a particular class of transformations called residual gauge transformations \cite{lenz}. 

In the following subsection we identify the states of the system as Schr\"odinger wave functionals and determine the general form of a physical state using the ``coordinate" representation for the canonical operators of the Maxwell field and the Gauss law constraint. From there we identify a general unitary gauge fixing transformation $U_g$ as a map from the physical space of states $\mathcal{H}_p$ to a space $\mathcal{H}_g$, which is the space of states for the gauge $g$. Next we determine the effect of this transformation on the various operators of the theory and express the Hamiltonian in the arbitrary gauge $g$. We conclude by using the Hamiltonian to calculate the Dirac equation in the gauge $g$.

\subsection{Gauge in classical electrodynamics}\label{2.1}

In electrodynamics the electric and magnetic fields defined by 
\begin{eqnarray}\label{eb}
{\bf E}&=&-\nabla\phi-{\partial {\bf A} \over \partial t}, \nonumber \\ {\bf B}&=&\nabla \times {\bf A} 
\end{eqnarray}
are invariant under a change of gauge;
\begin{eqnarray}\label{gt}
{\bf A}'&=&{\bf A}+\nabla f, \nonumber \\\phi'&=&\phi-{\partial f \over \partial t}
\end{eqnarray}
for an arbitrary function $f$. Clearly, however, the transverse vector potential ${\bf A}_{\rm T}$ is gauge invariant, while the (redundant) gauge dependent degrees of freedom are  longitudinal. For this reason the transverse vector potential serves as a convenient starting point relative to which vector potentials in other gauges can be defined by
\begin{eqnarray}\label{a}
{\bf A}={\bf A}_{\rm T}+\nabla f \,
\end{eqnarray}
with the function $f$ determining the gauge. The gauge invariant transverse electric field is given by
\begin{eqnarray}\label{et}
{\bf E}_{\rm T} =-{\partial {\bf A}_{\rm T}\over \partial t}\, 
\end{eqnarray}
while Gauss' law 
\begin{eqnarray}
\nabla \cdot {\bf E} = \rho
\end{eqnarray}
involving the charge density $\rho$ ensures that the longitudinal electric field is equal to minus the gradient of the (static) Coulomb potential; ${\bf E}_{\rm L} = -\nabla V$. Using this equality and eq.(\ref{eb}) we see that the scalar potential accompanying the vector potential in eq.(\ref{a}) can be written
\begin{eqnarray}\label{phi}
\phi = V - {\partial f \over \partial t} \, .
\end{eqnarray}

The Coulomb gauge is defined by the choice $f \equiv 0$, but it is not the only way in which the longitudinal degrees of freedom can be eliminated. In non-relativistic QED it has been shown \cite{woolley1} that a general gauge fixing condition is given by a linear functional constraint satisfied by vector potentials of the form
\begin{eqnarray}\label{a2}
{\bf A}={\bf A}_{\rm T}+\nabla\int {\rm d}^3x' {\bm g}({\bf x}',{\bf x})\cdot{\bf A}_{\rm T}({\bf x}') \,
\end{eqnarray}
where $ {\bm g}({\bf x},{\bf x}')$ is the Green function for the divergence operator;
\begin{eqnarray}\label{g}
\nabla\cdot  {\bm g}({\bf x},{\bf x}') = \delta({\bf x}-{\bf x}')\, .
\end{eqnarray}
In eq.({\ref{a2}) the redundant degrees of freedom have been re-expressed through a functional of the gauge invariant transverse degrees of freedom. 

While the longitudinal component of the Green function ${\bm g}$ is fixed according to eq.(\ref{g}) by
\begin{eqnarray}\label{gL}
{\bm g}_{\rm L}({\bf x},{\bf x}') = -\nabla \frac{1}{ 4\pi \vert {\bf x} - {\bf x}'\vert},
\end{eqnarray}
its transverse component is essentially arbitrary meaning that the gauge is determined through a choice of ${\bm g}_{\rm T}$. This idea has been employed in non-relativistic QED to obtain a Hamiltonian in an arbitrary gauge \cite{woolley1}, but it has yet to be extended to the relativistic setting. Furthermore the nature of gauge transformations and residual symmetries in such a framework have not been explored.

\subsection{The QED Lagrangian and Hamiltonian}

We start formally with the QED Lagrangian density
\begin{eqnarray}\label{L}
\mathscr{L} = {\rm i}\gamma_0 \gamma^\mu \psi^\dagger D_\mu \psi - (\gamma_0 m + e\phi_e)\psi^\dagger \psi-{1\over 4}F_{\mu\nu}F^{\mu\nu}
\end{eqnarray}
where $D_\mu = \partial_\mu +{\rm i}eA_\mu$ is the gauge covariant derivative, $F_{\mu\nu} = \partial_\mu A_\nu - \partial_\nu A_\mu$ is the electromagnetic field strength tensor and $\phi_e$ is an external potential due, for example, to nuclei.

Since the Lagrangian is independent of the velocity of the scalar potential its conjugate momentum is identically zero. As a result it is natural to quantize the theory within the Weyl gauge corresponding to the choice $\phi \equiv 0$. The remaining redundant degrees of freedom are eliminated by defining the physical subspace of states $\mathcal{H}_p$ consisting of those states, which vanish under the action of the Gauss law constraint; $G\ket{\varphi_p} \equiv (\nabla\cdot{\bf E}-\rho)\ket{\varphi_p} = 0$.

The Hamiltonian density is obtained from the Lagrangian density via a Legendre transformation;
\begin{eqnarray}\label{H}
\mathscr{H} = -{\rm i}\psi^\dagger {\bm \alpha}\cdot(\nabla-{\rm i}e{\bf A})\psi + (\beta m + e\phi_e) \psi^\dagger \psi + {1\over 2}\big({\bf \Pi}^2+(\nabla \times {\bf A})^2\big)\, .
\end{eqnarray}
Quantum mechanically $\psi$ and its conjugate $\psi^\dagger$ are Dirac field operators satisfying the anti-commutation relation
\begin{eqnarray}\label{com1}
\{\psi_\alpha({\bf x}),\psi^\dagger_\beta({\bf x}')\}=\delta_{\alpha\beta}\delta({\bf x}-{\bf x}'),
\end{eqnarray}
while ${\bf A}$ and ${\bf \Pi}=-{\bf E}$ are the canonical operators of the Maxwell field satisfying the commutation relation
\begin{eqnarray}\label{com2}
[A_i({\bf x}),{\rm \Pi}_j({\bf x}')]={\rm i}\delta_{ij}\delta({\bf x}-{\bf x}')\, .
\end{eqnarray}

We employ the usual definitions of the charge and current densities 
\begin{eqnarray}\label{rhoj}
\rho &=& e\psi^\dagger \psi, \nonumber \\ {\bf j} &=& e\psi^\dagger {\bm \alpha}\psi,
\end{eqnarray}
in terms of which the (conserved) Noether 4-current is $j^\mu = (\rho,{\bf j})$. Related to the charge density is the polarization field ${\bf P}_g$ defined by 
\begin{eqnarray}\label{Pg}
{\bf P}_g({\bf x}) = -\int {\rm d}^3x'\, {\bm g}_({\bf x},{\bf x}')\rho({\bf x}')
\end{eqnarray}
whose longitudinal component satisfies
\begin{eqnarray}\label{P2}
-\nabla \cdot {\bf P}_g = \rho\, ,
\end{eqnarray}
but whose transverse component is arbitrarily determined by ${\bm g}_{\rm T}$. Eq.(\ref{P2}) bares close resemblance to Gauss' law $G$, which can indeed be written
\begin{eqnarray}\label{G}
G= \nabla \cdot {\bf \Pi} + \rho = \nabla \cdot ({\bf \Pi}-{\bf P}_g).
\end{eqnarray}

As an operator $G$ is a symmetry of the Hamiltonian; $[G,H]=0$, and is responsible for generating time-independent gauge transformations of the vector potential and Dirac field operators. Identifying a group $(\{\beta({\bf x})\},+)$ consisting of real valued functions on $\mathbb{R}^3$ and group operation of addition, we define a group action $\Phi$ acting on the vector potential and Dirac fields by
\begin{eqnarray}\label{grpact}
\Phi[\psi, \beta] &=& e^{-{\rm i}e\beta}\psi, \nonumber \\ \Phi[{\bf A}, \beta] &=& {\bf A}+\nabla\beta\, .
\end{eqnarray}
The action is implemented through unitary transformations $\Omega$ generated by $G$, viz. $\Omega \psi \Omega^{-1} = \Phi[\psi,\beta] $ and $\Omega{\bf A}\Omega^{-1} = \Phi[{\bf A},\beta] $ where
\begin{eqnarray}\label{resgt}
\Omega[\beta] = \exp \bigg({\rm i}\int {\rm d}^3x \, ({\bf \Pi}\cdot \nabla + \rho)\beta({\bf x}) \bigg)\, .
\end{eqnarray}
These transformations are called {\em residual} gauge transformations, with the word residual intended to signify that the above time independent symmetry is what remains of the local gauge symmetry present in the original formulation \cite{lenz}.

\subsection{Unitary gauge fixing transformations}

We now turn our attention to the procedure of gauge fixing. In order to determine the form of a general gauge fixing transformation we first need to identify the form of a physical state. To do this we take as a Hilbert space ${\mathcal H}$ for the composite system wave functionals $\varphi[{\bf A}]$ of the $c$-number vector potential ${\bf A}$, taking values in the Hilbert space of the Dirac field operators \cite{jackiw}. 

A realization of the algebra of the Maxwell field operators ${\bf A}$ and ${\bf \Pi}$ is given on ${\mathcal H}$ using the ``coordinate" representation
\begin{eqnarray}\label{oprep}
({\hat{\bf A}}\varphi)[{\bf A}] &=& {\bf A}\varphi[{\bf A}] \nonumber \\ ({\hat {\bf \Pi}}\varphi)[{\bf A}] &=& -{\rm i}{\delta \varphi[{\bf A}] \over \delta {\bf A}} \, 
\end{eqnarray}
where we have introduced hats to distinguish between operators and $c$-number vector fields. Defining a scalar function $\alpha$ by $\nabla \alpha = {\bf A}_{\rm L}$, we can vary the wave functional $\varphi$ with respect to $\alpha$ and make use of eq.(\ref{oprep}) to obtain
\begin{eqnarray}\label{st}
{\rm i}{\delta \varphi \over \delta \alpha} = \nabla \cdot {\hat {\bf \Pi}}\varphi\, .
\end{eqnarray}
Using the constraint $G$ in eq.(\ref{G}) we get for a physical state $\varphi_p$
\begin{eqnarray}\label{physst}
{\rm i}{\delta \varphi_p \over \delta \alpha} = -\rho\varphi_p\,
\end{eqnarray}
and finally solving this equation gives the general form of a physical state;
\begin{eqnarray}\label{physst2}
\varphi_p[{\bf A}]=\varphi_p[{\bf A}_{\rm T}+\nabla\alpha]= \exp\bigg({\rm i}\int {\rm d}^3x \, \alpha({\bf x})\rho({\bf x}) \bigg)\varphi_p[{\bf A}_{\rm T}] \, .
\end{eqnarray}

Having determined the form of a physical state we can begin to define some unitary gauge fixing transformations. In the original work of Lenz et.al \cite{lenz} a unitary gauge fixing transformation yielding the Coulomb gauge representation was given as 
\begin{eqnarray}\label{ucoul}
U \equiv \exp\bigg(-{\rm i}\int {\rm d}^3x \, {\hat \alpha}({\bf x})\rho({\bf x}) \bigg) \,
\end{eqnarray}
where ${\hat \alpha}$ is defined analogously to $\alpha$ by $\nabla {\hat \alpha} = {\hat {\bf A}}_{\rm L}$. In the present context we see clearly that $U$ eliminates the dependence of the physical state on ${\bf A}_{\rm L}$;
\begin{eqnarray}\label{ucoul2}
(U\varphi_p)[{\bf A}] = \varphi_p[{\bf A}_{\rm T}]\, .
\end{eqnarray}

Now, as in eq.(\ref{a2}) of section \ref{2.1}, we write the longitudinal vector potential as the gradient of a functional of the transverse vector potential;
\begin{eqnarray}\label{a3}
{\bf A}_{\rm L}= \nabla \chi_g({\bf x},[{\bf A}_{\rm T}])
\end{eqnarray}
where
\begin{eqnarray}\label{chi}
\chi_g({\bf x},[{\bf A}_{\rm T}]) = \int {\rm d}^3x' {\bm g}({\bf x}',{\bf x})\cdot{\bf A}_{\rm T}({\bf x}') \, .
\end{eqnarray}
We can then define a more general unitary gauge fixing transformation $U_g$ by
\begin{eqnarray}\label{ug}
U_g \equiv \exp\bigg(-{\rm i}\int {\rm d}^3x \, \big({\hat \alpha}({\bf x})-\chi_g({\bf x},[{\hat {\bf A}}_{\rm T}])\big)\rho({\bf x}) \bigg),
\end{eqnarray}
mapping from $\mathcal{H}_p$ to an isomorphic space denoted $\mathcal{H}_g$, which is the space of states for the gauge $g$;
\begin{eqnarray}\label{ug2}
(U_g\varphi_p)[{\bf A}] &=& \exp\bigg({\rm i}\int {\rm d}^3x\,\chi_g({\bf x},[{\bf A}_{\rm T}])\rho({\bf x})\bigg)\varphi_p[{\bf A}_{\rm T}] \nonumber \\ &=& \varphi_p[{\bf A}_{\rm T}+\nabla\chi_g] = \varphi_p[{\bf A}] \equiv\varphi_g[{\bf A}_{\rm T}] \in {\mathcal H}_g
\end{eqnarray}
where ${\bf A} \equiv {\bf A}_{\rm T} + \nabla\chi_g$. The (transverse component of the) Green function ${\bm g}$ is essentially arbitrary and determines the gauge. Two commonly used examples are the Coulomb gauge; ${\bm g}_{\rm T}\equiv 0$ and the Poincar\'{e} gauge; $g_{{\rm T},j}({\bf x}',{\bf x}) \equiv -\int_0^1 {\rm d}\lambda x_i \delta_{ij}^{\rm T}({\bf x}'-\lambda{\bf x})$ \cite{power1,cohen1}.

The vector potential operator in the gauge $g$ is ${\hat {\bf A}}({\bf x}) \equiv {\hat {\bf A}}_{\rm T}({\bf x}) + \nabla\chi_g({\bf x},[ {\hat {\bf A}}_{\rm T}])$ with action on ${\mathcal H}_g$ given by
\begin{eqnarray}\label{Ag}
({\hat {\bf A}}\varphi_p)[{\bf A}] = ({\bf A}_{\rm T}+\nabla\chi_g)\varphi_g[{\bf A}_{\rm T}] \equiv ({\hat {\bf A}}\varphi_g)[{\bf A}_{\rm T}]\, .
\end{eqnarray}

Finally we define a Unitary transformation from a fixed gauge $g$ to a fixed gauge $g'$ by
\begin{eqnarray}\label{ugg'}
U_{gg'} \equiv \exp\bigg(-{\rm i}\int {\rm d}^3x \, \big(\chi_g({\bf x},[{\hat {\bf A}}_{\rm T}])-\chi_{g'}({\bf x},[{\hat {\bf A}}_{\rm T}])\big)\rho({\bf x}) \bigg)\,
\end{eqnarray}
an example of which is the well known Power-Zienau-Woolley transformation \cite{power1,craig,cohen1} used to obtain the Hamiltonian in the Poincar\'{e} gauge from the Hamiltonian in the Coulomb gauge. Such a gauge transformation is not to be confused with the residual gauge (symmetry) transformation given in eq.(\ref{resgt}).

\subsection{The Hamiltonian in the gauge $g$}

To obtain the Hamiltonian in the gauge $g$ we need to determine the effect of the transformation in eq.(\ref{ug}) on the various operators of the theory, namely $\psi,\, \psi^\dagger, \, {\bf A}$ and ${\bf \Pi}$. In doing so we will resume denoting operators without hats. Clearly $U_g$ leaves the vector potential ${\bf A}$ unchanged, while the action of ${\bf A}$ on ${\mathcal H}_g$ is given in eq.(\ref{Ag}). The effect on the Dirac field operator $\psi$ is that of a gauge transformation
\begin{eqnarray}\label{t1}
U_g\psi U_g^{-1} = e^{{\rm i}e(\alpha - \chi_g)}\psi\, .
\end{eqnarray}
The canonical momentum ${\bf \Pi}$ transforms as
\begin{eqnarray}\label{t2}
U_g{\bf \Pi} U_g^{-1} = {\bf \Pi} + {\bf P}_g
\end{eqnarray}
so that in the new representation ${\bf \Pi}$ represents the (negative of) the gauge dependent displacement operator ${\bf D}_g\equiv {\bf E}+{\bf P}_g$. Using eq.(\ref{P2}) we find the constraint $G$ and the residual gauge transformation $\Omega$ transform as follows
\begin{eqnarray}
U_gGU_g^{-1} &=& \nabla\cdot{\bf \Pi}, \nonumber \\ U_g\Omega[\beta] U_g^{-1} &=& \exp\bigg({\rm i}\int {\rm d}^3x\, ({\bf \Pi}\cdot\nabla)\beta({\bf x}) \bigg),
\end{eqnarray}
which are independent of the gauge $g$. The constraint $G$ implies that the longitudinal canonical momentum ${\bf \Pi}_{\rm L}$ vanishes on $\mathcal{H}_g$. On the one hand this means ${\bf P}_{\rm L}$ alone represents (the negative of) the longitudinal electric field, and on the other that the Hamiltonian density on $\mathcal{H}_g$ can be written in terms of the transverse operators ${\bf A}_{\rm T}$ and ${\bf \Pi}_{\rm T}$ only;
\begin{eqnarray}\label{h2}
\mathscr{H} &=& -{\rm i}\psi^\dagger {\bm \alpha}\cdot(\nabla-{\rm i}e({\bf A}_{\rm T}+\nabla\chi_g))\psi + (\beta m + e\phi_e) \psi^\dagger \psi \nonumber \\ && +\,  {1\over 2}{{\bf P}_{\rm L}}^2 + {1\over 2}\big(({\bf \Pi}_{\rm T} + {\bf P}^g_{\rm T})^2+(\nabla \times {\bf A}_{\rm T})^2\big)\, 
\end{eqnarray}
where ${\bf A}_{\rm T}+\nabla\chi_g$ is simply the vector potential ${\bf A}$ in the gauge $g$. Eq.(\ref{h2}) gives a Hamiltonian in an arbitrary gauge, which is fully relativistic in the material degrees of freedom. It is one of the main results of this paper.

The commutator of the transverse operators follows from eq.(\ref{com2}) and is given by
\begin{eqnarray}\label{com3}
[A_{\rm T,i}({\bf x}),{\rm \Pi}_{\rm T,j}({\bf x}')]={\rm i}\delta^{\rm T}_{ij}({\bf x}-{\bf x}')
\end{eqnarray}
with $\delta^{\rm T}$ denoting the transverse delta function. We note also that denoting the Fourier transforms of  ${\bf A}_{\rm T}$ and ${\bf \Pi}_{\rm T}$ with tildes we can define photon creation and annihilation operators in the usual way;
\begin{eqnarray}\label{phot}
a_\lambda ({\bf k}) = \sqrt{1\over 2\omega}\bigg(\omega {\tilde A}_{\rm T,\lambda}({\bf k})+{\rm i}{\tilde \Pi}_{\rm T,\lambda}({\bf k})\bigg)
\end{eqnarray}
where $\lambda = 1,2$ denotes one of two polarization directions orthogonal to ${\bf k}$. The bosonic commutator 
\begin{eqnarray}\label{com4}
[a_\lambda({\bf k}),a_{\lambda'}^\dagger({\bf k}')] = \delta_{\lambda\lambda'}\delta({\bf k}-{\bf k}'),
\end{eqnarray}
follows from eq.(\ref{com3}).

\subsection{The Dirac equation in the gauge $g$}

It is an instructive exercise to calculate in the arbitrary gauge $g$, the equation of motion for the Dirac field operator $\psi$, which should be the Dirac equation in the presence of a Maxwell field. The calculation demonstrates how the scalar potential, like the longitudinal vector potential is re-expressed through the functional $\chi_g$. 

Writing first the products of Dirac field operators appearing in eq.(\ref{h2}) in normal order, we obtain
\begin{eqnarray}\label{d1}
{\rm i}{\dot \psi} &=& \bigg[ {\bm \alpha}\cdot \big(-{\rm i}\nabla-e({\bf A}_{\rm T}+\nabla\chi_g)\big) +\beta m+e\phi_e \nonumber \\ && + \, {e\over 4\pi}\int {\rm d}^3x'\, {\rho({\bf x}')\over \vert {\bf x} - {\bf x}' \vert} - e\int {\rm d}^3x'\,{\bm g}_{\rm T}({\bf x},{\bf x}')\cdot \big({\bf \Pi}_{\rm T} + {\bf P}_g^{\rm T}\big) \bigg]\psi \, .
\end{eqnarray}
The first term on the second line of eq.(\ref{d1}) is equal to $eV$ with $V$ denoting the static Coulomb potential of charges. The transverse electric field in the gauge $g$ is ${\bf E}_{\rm T}= -({\bf \Pi}_{\rm T} + {\bf P}_g^{\rm T})$ and it is straightforward to verify that ${\bf E}_{\rm T}=-{\dot {\bf A}}_{\rm T}$, as in eq.(\ref{et}). These equalities imply that eq.(\ref{d1}) can be written
\begin{eqnarray}\label{d2}
{\rm i}{\dot \psi} = \bigg[ {\bm \alpha}\cdot \big(-{\rm i}\nabla-e{\bf A}\big) +\beta m+e(\phi_e +\phi) \bigg]\psi \, .
\end{eqnarray}
where we have defined the scalar potential anew by
\begin{eqnarray}\label{phi2}
\phi = V - {\partial \chi_g \over \partial t},
\end{eqnarray}
which is analogous to eq.(\ref{phi}).

\section{Some implications of the formalism}

Having obtained the Hamiltonian in an arbitrary gauge we discuss in this section some implications resulting from the freedom to choose ${\bm g}_{\rm T}$. We first point out that as in the non-relativistic case a canonical partitioning of the Hamiltonian is gauge dependent and therefore leads to gauge dependent definitions of quantum subsystems \cite{woolley1}.  

By means of analogy with classical electrodynamics, we go on to demonstrate how gauge invariant subsystems components might be defined. Finally we consider an application of such a definition in the context of energy transfer and causality in Fermi's two atom problem.

\subsection{Ambiguity in defining quantum systems}

To understand what effect the arbitrariness of ${\bm g}_{\rm T}$ might have it is important to identify the physical observables in a given gauge $g$. In the Weyl gauge the canonical momentum ${\bf \Pi}$ is the negative of the electric field ${\bf E}$. In the gauge $g$ we have ${\bf E}=-U_g{\bf \Pi}U_g^{-1} = -({\bf \Pi}+{\bf P}_g)$. The operator ${\bf \Pi}_{\rm T}+{\bf P}_{\rm T}^g$ appearing in eq.(\ref{h2}) therefore represents the negative of the transverse electric field ${\bf E}_{\rm T}$. This identity was used in obtaining the Dirac equation eq.(\ref{d2}). 

Due to the gauge dependence of ${\bf P}_{\rm T}^g$, the operator ${\bf \Pi}_{\rm T}$ is implicitly gauge dependent in that it represents a different physical observable in each different gauge. Explicitly ${\bf \Pi}_{\rm T}$ represents the transverse component of (the negative of) the gauge dependent Displacement operator ${\bf D}_g$. 

Now, the most common way to use a Hamiltonian (density) such as eq.(\ref{h2}) is to split it into ``free" and ``interacting" components as follows;
\begin{eqnarray}\label{h0,I}
\mathscr{H} &=& \mathscr{H}_0+\mathscr{H}_{\rm I}
\end{eqnarray}
where
\begin{eqnarray}\label{h0}
\mathscr{H}_0 &\equiv & \mathscr{H}_{\rm D}+\mathscr{H}_{\rm EM}, \nonumber \\
\mathscr{H}_{\rm D} &\equiv & 
-{\rm i}\psi^\dagger {\bm \alpha}\cdot\nabla\psi + (\beta m + e\phi_e) \psi^\dagger \psi + {1\over 2}{\bf P}_{\rm L}^2 + {1\over 2}{{\bf P}_{\rm T}^g}^2, \nonumber \\
\mathscr{H}_{\rm M} &\equiv & {1\over 2}\big({\bf \Pi}_{\rm T}^2+(\nabla \times {\bf A}_{\rm T})^2\big)
\end{eqnarray}}
and
\begin{eqnarray}\label{hI}
\mathscr{H}_{\rm I} &\equiv & -e\psi^\dagger {\bm \alpha}\cdot({\bf A}_{\rm T}+\nabla\chi_g)\psi + {\bf \Pi}_{\rm T}\cdot{\bf P}_{\rm T}^g\, .
\end{eqnarray}
The reason for this splitting is of course, that the sets of operators $\{\psi,\psi^\dagger\}$ and $\{{\bf A}_{\rm T},{\bf \Pi}_{\rm T}\}$ are mutually commuting. The component $
\mathscr{H}_{\rm D}$ represents the Dirac field ``subsystem", $\mathscr{H}_{\rm M}$ the Maxwell field subsystem and $\mathscr{H}_{\rm I}$ their interaction. The problem with such a splitting is that the subsystem components so defined are like the operator ${\bf \Pi}_{\rm T}$, implicitly gauge dependent and as a result physically ambiguous. 

Given this ambiguity it is natural to try and determine what kind of calculations can be carried out, that yield results independent of $g$ i.e. that are gauge invariant. For example, $S$-matrix elements on energy shell are gauge invariant to all orders in perturbation theory \cite{woolley3,woolley2}. Consequently the conceptual difficulty regarding the definition of subsystems does not effect scattering theory in any way. The invariance of $S$-matrix elements rests on the fact that the bare states (eigenstates of the free energy operator $H_0$) coincide asymptotically with eigenstates of the total Hamiltonian, that is, that bare states are asymptotically stable.

In order that conventional (perturbative) calculations using bare states produce gauge invariant results a condition of free energy conservation must be satisfied \cite{woolley3}. Otherwise calculations will in general yield gauge dependent results. Free energy conservation is a condition which must be imposed from outside the theory, so various approximations, which essentially ensure free energy conservation by giving rise to a delta function in the initial energy minus the final energy of the process under study, are used throughout non-relativistic QED and quantum optics; they include the resonant state on energy shell approximation \cite{blake,woolley3}, the pole approximation \cite{barnett}, the Markovian approximation \cite{barnett} and the Fermi approximations \cite{craig,cohen2}.

A general survey of the types of calculation used in practice in non-relativistic QED and their dependence on $g$ can be found in \cite{woolley3}. A typical example of the difference in predicted results from calculations in two different gauges is well known in non-relativistic QED where the Coulomb gauge (minimal coupling) Hamiltonian and Poincar\'{e} gauge (multipolar) Hamiltonian yield different results for, among other things, the theoretical lineshape of spontaneous emission \cite{power2}. The source of this difference lies in the use of physically different canonical operators in determining the lineshape observable.

The ambiguity regarding subsystems defined using canonical variables is also present at the classical level. It occurs when moving to the Hamiltonian formalism from the Lagrangian formalism. Since canonical momenta are defined in terms of the Lagrangian, equivalent Lagrangians yielding equivalent Hamiltonians will in general not yield physically equivalent canonical momenta. Again, a well known example is given by the Coulomb gauge and Poincar\'{e} gauge formulations of classical electrodynamics.

At the quantum level the ambiguity in the definition of subsystems can be viewed as a generic trait of interacting theories, whether they are relativistic, field-theoretic or otherwise. Given a Hamiltonian dependent on two sets of mutually commuting operators $\{x_i\}$ and $\{y_i\}$ and a splitting of the Hamiltonian of the form $H=H_x(x_i)+H_y(y_i)+H_{xy}(x_i,y_i)$, an equivalent Hamiltonian is obtained by a unitary transformation $H'=UHU^{-1}$. In general, the subsystem components of $H'$ will not be equivalent to those of $H$ i.e. $H'_x \neq UH_xU^{-1}$, with the same being true for $H_y$ and $H_{xy}$. The importance of this fact for concepts such as quantum entanglement and decoherence has been recognized in the philosophy literature \cite{dugic}.

\subsection{``Classical-type" gauge invariant subsystems and operators}

In the previous subsection we discussed the gauge dependence of splitting the Hamiltonian into ``free" and ``interaction" components and we reviewed the situation regarding the kinds of calculation, that yield gauge invariant results. 

In this section we address the complimentary question as to whether {\em manifestly} gauge invariant subsystem components can be defined from the outset. This question has, so far, received no direct attention in the literature, but it is important if one wishes to identify how gauge invariant results might be obtained outside the range of validity of the simplifying assumptions of scattering theory.

To see how we might define gauge invariant subsystems consider first the case of a free classical electron coupled to a classical Maxwell field. For this system the Hamiltonian in {\em any} gauge can be written
\begin{eqnarray}\label{hclas}
H &=& {1 \over 2} m{\dot {\bf r}}^2 + \frac{1}{2} \int {\rm d}^3 {\bf{x}} \, \bigg({\bf E}({\bf x})^2 + {\bf B}({\bf x})^2\bigg),
\end{eqnarray}
which represents the total energy of the system as the sum of the kinetic energy of the electron and the energy of the electromagnetic (EM) field. Regarding the electron variables what varies between gauges is the identification of the gauge invariant velocity ${\dot {\bf r}}$, with the electron canonical momentum ${\bf p} = ({\dot {\bf r}}+e{\bf A})/m$ \cite{craig,woolley1,power3}.

The classical electron velocity and canonical momentum have clear analogues in QED; the ``velocity density"
\begin{eqnarray}\label{v}
{\mathscr V} = -{\rm i}\psi^\dagger {\bm \alpha}\cdot(\nabla-{\rm i}e({\bf A}_{\rm T}+\nabla\chi_g))\psi
\end{eqnarray}
is manifestly gauge invariant, while the canonical momentum density ${\mathscr P}=-{\rm i}\psi^\dagger {\bm \alpha}\cdot\nabla\psi$ depends on $g$. Analogously to eq.(\ref{hclas}) we can split the Hamiltonian into two gauge invariant components
\begin{eqnarray}\label{h4}
\mathscr{H} &=& \mathscr{H}_{\rm M} + \mathscr{H}_{\rm EM}, \nonumber \\
\mathscr{H}_{\rm M} &=& -{\rm i}\psi^\dagger {\bm \alpha}\cdot(\nabla-{\rm i}e({\bf A}_{\rm T}+\nabla\chi_g))\psi + (\beta m + e\phi_e) \psi^\dagger \psi \nonumber \\ 
\mathscr{H}_{\rm EM} &=&  {1\over 2}{{\bf P}_{\rm L}}^2 + {1\over 2}\big(({\bf \Pi}_{\rm T} + {\bf P}^g_{\rm T})^2+(\nabla \times {\bf A}_{\rm T})^2\big)
\end{eqnarray}
with the first component representing the energy density of the matter field and the second component the energy density of the EM field. With these new definitions the Hamiltonian naturally represents the energy of the system as the sum of energies of the subsystems, rather than as the sum of superficially defined free and interaction energies. 

We note that the first term in $\mathscr{H}_{\rm EM}$ represents the energy density of the longitudinal EM field, while the second represents the energy density of the transverse EM field. The subsystems defined in this way are coupled, because the velocity density ${\mathscr V}$ and the electric field ${\bf E}$ do not commute; using for simplicity, the Coulomb gauge (${\bm g}_{\rm T} \equiv 0$), we obtain
\begin{eqnarray}\label{vE}
[{\mathscr V}({\bf x}),{\rm E}_j({\bf x}')] &=& [{\rm i}\psi^\dagger({\bf x}) {\bm \alpha}\cdot(\nabla-{\rm i}e{\bf A}_{\rm T}({\bf x}) )\psi({\bf x}) ,{\rm \Pi}_{{\rm T},j}({\bf x}') +{\rm P}_{{\rm L},j}({\bf x}') ] \nonumber \\ 
&=& {\rm i}e\psi^\dagger({\bf x}) \alpha_i \bigg(\delta_{ij}^{\rm T}({\bf x}-{\bf x}')-\nabla_i \nabla_j {1 \over 4\pi \vert {\bf x}-{\bf x}'\vert} \bigg)\psi({\bf x})  \nonumber \\ &\equiv & {\rm i}e\psi^\dagger({\bf x}) \alpha_i \delta_{ij}\delta({\bf x}-{\bf x}')\psi({\bf x})  \nonumber \\ &\equiv & {\rm i}{\rm j}_j({\bf x})\delta({\bf x}-{\bf x}') \, .
\end{eqnarray}
The delta function ensures that ${\mathscr V}({\bf x})$ and ${\bf E}({\bf x}')$ are compatible observables for ${\bf x}\neq{\bf x}'$. Moreover it ensures that the matter field and the EM field energies are compatible in disjoint regions i.e.
\begin{eqnarray}\label{hcom}
[H^{\mathcal{R}}_{\rm M},H^{\mathcal{R}'}_{\rm EM}]\equiv \int_{\mathcal{R}} {\rm d}^3x \int_{\mathcal{R}'} {\rm d}^3x'\, [\mathscr{H}_{\rm M}({\bf x}),\mathscr{H}_{\rm EM}({\bf x}')] =0
\end{eqnarray}
whenever ${\mathcal R}\cap{\mathcal R}' = \emptyset$.

Using the commutator eq.(\ref{vE}) and the following commutator of the electric field and magnetic field energy 
\begin{eqnarray}\label{ebcom}
{1\over 2} \int {\rm d}^3x' \, [{\rm E}_i({\bf x}), {\rm B}_j({\bf x})^2] = {\rm i}(\nabla \times {\rm B}({\bf x}))_i\, ,
\end{eqnarray}
the equation of motion for the electric field is found to be
\begin{eqnarray}\label{Edot}
{\dot {\bf E}}= \nabla \times {\bf B} -{\bf j},
\end{eqnarray}
which is just one of Maxwell's equations. The remaining Maxwell equation 
\begin{eqnarray}\label{Bdot}
{\dot {\bf B}}= -\nabla \times {\bf E}
\end{eqnarray}
is found in a similar fashion as the equation of motion for the magnetic field.

\subsection{Energy and Causality}

In recent years a large amount of attention has been given to the nature of energy transfer between separated material systems \cite{fermi,shirokov,hegerfeldt,power4,persico,milonni,buchholz,juanjo}. Specifically in the context of the two atom problem of Fermi \cite{fermi}. 

In order to investigate causality at the microscopic level, Fermi considered two identical spatially separated atoms $A$ and $B$. Initially atom $A$ is energetically excited while atom $B$ is in its ground state and there are no photons present in the EM field. The question posed by Fermi was; when does atom $B$ begin to move out of its ground state due to atom $A$? Einstein causality would appear to require that any changes in the energy of atom $B$ be independent of atom $A$ for all times less than the time it would take for a signal produced by atom $A$ travelling at the maximal speed of propagation $c$, to reach atom $B$. The most recent work concerning the Fermi problem was the proposal of a circuit QED experiment designed to test for any possible violations of causality \cite{juanjo}. 

The majority of theoretical ``proofs" of causality in the Fermi problem involve using the bare states of a non-relativistic Hamiltonian in the Poincar\'{e} gauge and the electric dipole approximation \cite{power4,persico,milonni,juanjo}. The electric dipole approximation dictates that the atoms in the Fermi problem couple to the Maxwell field at the $c$-number, atomic center of mass positions. The Fermi problem can then be formulated in terms of a well-defined, center of mass separation. Moreover the electric dipole approximation ensures that the dipole canonical momenta ${\rm p}_i = m{\dot {\rm r}}_i$, which define the bare atomic energies, are purely kinetic \cite{craig,cohen1}. 

At the same time in the Poincar\'{e} gauge the field canonical momentum is identified as the (negative of the) local multipolar transverse displacement field \cite{craig,cohen1}. Outside the atoms, which in the electric dipole approximation means away from the center of mass positions, this field coincides with the retarded electric field. The bare energies of the atoms are coupled through this field, which ensures that there are no violations of causality.

A different proof presented in \cite{buchholz} uses the abstract language of algebraic quantum field theory, relying quite generally on the primitively causal nature of relativistic quantum field theory resulting from the hyperbolicity of the relevant equations of motion.

Here, we will show using the gauge invariant definition of $\mathscr{H}_{\rm M}$ in eq.(\ref{h4}) that changes in the energy density of the matter field at a point $(t,{\bf x})$ are independent of the matter field at all points, which cannot be connected  to $(t,{\bf x})$ by a causal signal. The energy of the matter field in some closed region $\mathcal{R} \in {\mathbb R}^3$ is merely;
\begin{eqnarray}\label{hm}
H^{\mathcal{R}}_{\rm M}(t)&=& \int_{\mathcal{R}} {\rm d}^3x \, \mathscr{H}_{\rm M}({\bf x},t)\, .
\end{eqnarray}
Since $\mathscr{H}_{\rm M}$ is gauge invariant the result does not rely on the use of a particular gauge and avoids any approximations. 

We begin by calculating the equation of motion for $\mathscr{H}_{\rm M}$, which for simplicity is carried out in the Coulomb gauge;
\begin{eqnarray}\label{hmdot}
{\rm i} {\dot {\mathscr H}_{\rm M}}({\bf x}) &=& \nonumber \\ && \hspace*{-1.7cm} -{1\over 2} \int {\rm d}^3 x' \, [ e\psi^\dagger({\bf x})\alpha_i {\rm A}_{\rm T,i} ({\bf x}) \psi({\bf x}) , {\rm \Pi}_{\rm T,j}({\bf x}')^2]+{\rm i}[\psi^\dagger({\bf x})\alpha_i \nabla_i \psi({\bf x}),{\rm P}_{\rm L,j}({\bf x}')^2] \nonumber \\ && \hspace*{-1.7cm} = {\rm i}{\rm j}_i({\bf x})({\rm E}_{\rm T,i}({\bf x})+{\rm E}_{\rm L,i}({\bf x})) \nonumber \\ && \hspace*{-1.7cm} = {\rm i}{\rm j}_i({\bf x}){\rm E}_i({\bf x})\, .
\end{eqnarray}
Thus, together with the Maxwell equations eqs.(\ref{Edot}) and (\ref{Bdot}) we have a system of equations, which can be written
\begin{eqnarray}\label{sys}
{\dot {\mathscr H}_{\rm M}} ({\bf x},t) &=& {\bf j} ({\bf x},t) \cdot {\bf E} ({\bf x},t), \nonumber \\ {\bf j} ({\bf x},t) &=& \nabla \times {\bf B} ({\bf x},t) -{\dot {\bf E}} ({\bf x},t), \nonumber \\ \Box {\bf E}({\bf x},t) &=& -\nabla\rho({\bf x},t) - {\dot {\bf j}}({\bf x},t), \nonumber \\ \Box {\bf B}({\bf x},t) &=& \nabla \times {\bf j}({\bf x},t),
\end{eqnarray}
where $\Box$ is the d'Alembertian ${\partial^2 / \partial t^2}-\nabla^2$. 

Now, first we note that using the second equation the first can be written in terms of the electric and magnetic fields alone. Second we note that the remaining two equations are inhomogeneous wave equations for the (cartesian components of the) electric and magnetic fields with source terms ${\bm \nu}({\bf x},t) \equiv \nabla\rho({\bf x},t) + {\dot {\bf j}}({\bf x},t)$ and ${\bm \mu}({\bf x},t) \equiv -\nabla \times {\bf j}({\bf x},t)$ respectively. These equations are hyperbolic and have well known retarded solutions of the form 
\begin{eqnarray}\label{sys2}
{\bf E}({\bf x},t) &=& {\bf E}_0({\bf x},t) + {\bf E}_{\rm r}({\bf x},t), \nonumber \\
{\bf B}({\bf x},t) &=& {\bf B}_0({\bf x},t) + {\bf B}_{\rm r}({\bf x},t)
\end{eqnarray}
where $ {\bf E}_0$ and $ {\bf B}_0$ satisfy the homogeneous equations $\Box{\bf E}_0 =\Box {\bf B}_0 =0$, while $ {\bf E}_{\rm r}$ and ${\bf B}_{\rm r}$ depend respectively on the sources ${\bm \nu}({\bf x}',t_{\rm r})$ and ${\bm \mu}({\bf x}',t_{\rm r})$ at the retarded time $t_{\rm r} = t - |{\bf x}-{\bf x}'|$ \cite{morse,jackson}. 

Using these solutions we can conclude that as desired ${\dot {\mathscr H}_{\rm M}}({\bf x},t)$ at the point $(t,{\bf x})$ depends on the matter field at points $(t_{\rm r},{\bf x'})$ only. 

\section{Conclusions}

In this paper we have derived a Hamiltonian in an arbitrary gauge, which is appropriate for the relativistic description of atoms and molecules. It also serves well as a means by which the gauge freedom of QED in Hamiltonian form can be explored. We have discussed some implications of this gauge freedom highlighting that a canonical partitioning of the Hamiltonian is manifestly gauge dependent. 

We have pointed out that because of this, how the calculation of gauge invariant results can be achieved is a non-trivial question. We have shown that a classical-type partitioning of the Hamiltonian in terms of ``velocities" instead of canonical operators is gauge invariant. Finally we have suggested a possible application of such a partitioning in relation to the problem of causality in spatially separated material systems.

\end{document}